\documentclass[twocolumn]{aastex631}
\usepackage{natbib, amsmath, float}
\usepackage{pifont}

\shorttitle{The CGM of isolated galaxies in eFEDS}
\shortauthors{Chadayammuri et al.}

\begin{document}

\title{Testing galaxy feedback models with resolved X-ray profiles of the hot circumgalactic medium}

\author{Urmila Chadayammuri}
\affiliation{Center for Astrophysics \ding{120} Harvard \& Smithsonian, 60 Garden St, Cambridge, MA - 02138, USA}

\author{\'Akos Bogd\'an}
\affiliation{Center for Astrophysics \ding{120} Harvard \& Smithsonian, 60 Garden St, Cambridge, MA - 02138, USA}

\author{Benjamin D. Oppenheimer}
\affiliation{Center for Astrophysics and Space Astronomy, 389 UCB, Boulder, CO - 80309, USA}

\author{Ralph P. Kraft}
\affiliation{Center for Astrophysics \ding{120} Harvard \& Smithsonian, 60 Garden St, Cambridge, MA - 02138, USA}

\author{William R. Forman}
\affiliation{Center for Astrophysics \ding{120} Harvard \& Smithsonian, 60 Garden St, Cambridge, MA - 02138, USA}

\author{Christine Jones}
\affiliation{Center for Astrophysics \ding{120} Harvard \& Smithsonian, 60 Garden St, Cambridge, MA - 02138, USA}

\begin{abstract}
The hot ($>10^6$ K) phase of the circumgalactic medium (CGM) contains a large fraction of baryons in galaxies. It also retains signatures of the processes that shaped the galaxies, such as feedback from active galactic nuclei (AGNs) and supernovae, and offers a uniquely powerful way to constrain theoretical models of feedback. It is, however, notoriously difficult to detect. By stacking 2643 optically selected galaxies in the eROSITA Final Equatorial Depth Survey (eFEDS), we present spatially resolved properties of the extended CGM in both star-forming and quiescent galaxies spanning an order of magnitude in stellar mass. We mask out resolved point sources and galaxy groups/clusters and model the contribution from X-ray binaries and the hot ISM, producing accurate radial profiles. We compare the profiles to mock X-ray observations of galaxy stacks in the IllustrisTNG100 (TNG) and EAGLE cosmological simulations. We detect extended emission from both the high-mass ($10.7<\log(M_*/M_\odot)<11.2$) and low-mass ($10.2<\log(M_*/M_\odot)<10.7$) galaxy stacks. Galaxies have somewhat more luminous CGM between $10-100$~kpc if they are more massive or star-forming. However, the luminosity increases slower with stellar mass than predicted in simulations. Simulated quenched galaxies are dimmer than observed, suggesting that they rely too heavily on CGM ejection for quenching. Star-forming galaxies are observed to have flatter and more extended profiles than in simulations, suggesting under-efficient stellar feedback models. Our results highlight the need to modify future prescriptions of galaxy feedback models.
\end{abstract}



\section{Introduction} 
\label{sec:main}
The CGM is defined as the gas that lives outside the galactic disk and within a galaxy's virial radius \citep[for a review, see][]{Tumlinson2017}. This atmosphere is partially accreted from the cosmic web, and reshaped and enriched over cosmic time by feedback from galactic processes \citep{Hafen2019}. It forms the reservoir of gas that fuels star formation over cosmic time, and retains the signatures of energetic feedback from stars, supernovae, and supermassive black holes. 

Most extensively, the CGM has been studied in the ultra-violet, notably with the Cosmic Origins Spectrograph (COS) on the Hubble Space Telescope, which measures the absorption of light from background quasars by the cool and warm phases of the CGM. These studies raised several questions -- where are all the baryons \citep{Stocke2013,Werk2014}? If there is abundant CGM in both star-forming and quiescent galaxies, what physical process quenches them \citep{Tumlinson2011,Borthakur2015,Thom2012}? Where are the metals that formed via stellar nucleosynthesis \citep{Peeples2014}? 

Theoretical models of galaxy formation predict the existence of a diffuse, hot phase  ($T>10^6$~K) that can generate line emission at X-ray wavelengths \citep{White1978,White1991}. This hot phase is expected to extend hundreds of kiloparsecs from the galaxy centers \citep{Werk2014,Oppenheimer2020,Davies2020}, contain 5-50$\%$ of the galaxy's baryons \citep{Stocke2013, Werk2016,Bregman2018}, and, due to its diffuse nature, retain unique signatures from different modes of energetic feedback \citep[][]{Liang2016,Jiang2018}. In simulations as diverse as IllustrisTNG \citep[][]{Truong2020}, EAGLE \citep{Davies2020}, and SIMBA \citep{robson2020}, star-forming galaxies are X-ray brighter than their quiescent counterparts since the quenching mechanism is the expulsion of gas from the star-forming galactic centers. Idealised simulations also find that stellar feedback produces a significant X-ray flux \citep{Sarkar2016,Fielding2020}.

The temperature and density of the hot CGM place it close to the detection threshold of current X-ray telescopes. \textit{Einstein} observatory first detected the hot coronae around the brightest and most nearby galaxies \citep[e.g.][]{Nulsen1984,Forman1985,Trinchieri1985}; \textit{Chandra} and XMM-\textit{Newton} have achieved higher sensitivity and much finer resolution \citep[e.g.][]{Anderson2011,Bogdan2013b,Bogdan2013a, Bogdan2015a,Bogdan2017,Anderson2016,Li2017,Li2018,Forbes2017,Das2019}. These studies detected extended and luminous X-ray coronae around massive spiral galaxies. \citet{Li2017} notes that the halos of more massive spirals seem to be in hydrostatic equilibrium,  like in ellipticals, whereas in lower-mass galaxies they seem to be hotter, likely due to stellar feedback dominating over gravity. Further, \cite{Mulchaey2010} found that elliptical galaxies in groups and clusters are much more luminous than field counterparts at fixed stellar mass. At least qualitatively, this picture agrees with that from current cosmological simulations. While informative, these conclusions are ultimately drawn from biased samples, and have limited ability to assess the effect of feedback in galaxies of different masses.

More representative samples have been obtained by stacking images from the ROSAT all-sky survey (RASS) around optically detected galaxies, which allows the signal to accumulate while the random noise cancels out on average. \citet{Bogdan2015} used $\sim3000$ galaxies from SDSS to measure the X-ray luminosity as a function of stellar mass. \citet{Anderson2013} stacked $\sim2100$ galaxies, selected from the 2MASS survey, in the ROSAT catalog to retrieve X-ray luminosity measurements within stacks of galaxies split by optical morphology; unlike the \textit{Chandra} and \textit{XMM}-Newton studies, they found elliptical galaxies to be over twice as luminous as spirals. Perhaps most ambitiously, \citet{Anderson2015} stacked ROSAT images around 250,000 ``locally brightest galaxies" (LBGs), measuring surface brightness profiles with a resolution of $\sim30$~kpc out to 1~Mpc. The latter two studies did not check whether the galaxies belonged to, or were along the line of sight to, galaxy groups or clusters, whose emission would dominate over the CGM. Further, 30~kpc is comparable to the full extent of the CGM in individually detected halos \cite{Li2018} and their simulated analogs \citep{Oppenheimer2020}, implying that ROSAT cannot resolve its internal structure.

In this paper, we build on previous population studies of the X-ray CGM, exploring its internal structure in stacks of isolated galaxies. This is made possible by the \textit{eROSITA} telescope, which  has $30-50$ times the sensitivity of ROSAT and two times finer angular resolution \citep{Merloni2020}. eROSITA Final Equatorial Depth Survey (eFEDS) is a 140~deg$\rm{^2}$ contiguous field with an almost uniform exposure of $\sim2.5$~ks, which is planned to be the all-sky exposure at the end of eight full-sky surveys. eFEDS has self-consistent group and cluster catalogs that allow us to eliminate contamination from any extraneous sources along the line of sight \citep{Brunner2021,Liu2021} in a way that was not feasible in earlier ROSAT studies. The sensitivity of \textit{eROSITA} allows us to split the sample into bins of stellar mass and then further into star-forming and quiescent galaxy classes. Moreover, the higher spatial resolution allows us to construct stacked X-ray surface brightness profiles of the CGM. We compare these profiles to analogous stacks from the TNG and EAGLE simulations \citep{Oppenheimer2020}. We discuss the differences, and their implications for how the galaxy evolution models in these state-of-the-art simulations need to evolve to match our evolving understanding of the CGM. We also compare our results to the recent, parallel work by \citet{Comparat2022}.

\section{Methods}

\subsection{Galaxy catalog}
 
\begin{figure*}
    \centering
    \includegraphics[width=\textwidth]{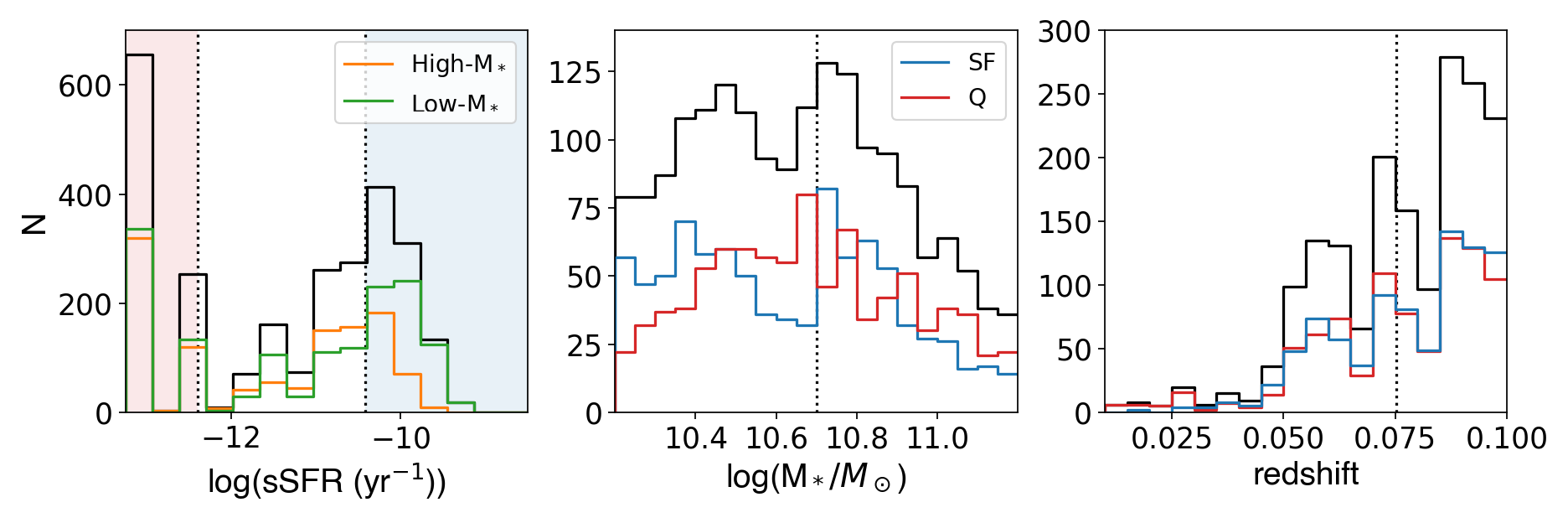}
    \caption{Distribution of the properties of the galaxies used in this study. We set a floor of 10$^{-13}$\ yr$^{-1}$ to the sSFR, since star-formation rates lower than $10^{-3} \ \rm{M_{\odot} \ yr^{-1}}$ would produce very weak signals. For each mass bin, the top third of galaxies by specific star formation rate (sSFR) make up the star-forming galaxy stack, while the bottom third are marked quiescent. This split is shown for the entire sample in the left panel. The dotted line in the second panel shows the division between the low-mass and high-mass samples. The blue and red histograms in the middle and right panels show the star-forming and quiescent galaxies, respectively.}
    \label{fig:sdss}
\end{figure*}
We produced the galaxy catalog from SDSS Data Release 16, which includes the RA, DEC, and spectroscopic redshift for each galaxy. We use stellar masses and star formation rates from the Granada catalog \citep{Montero2016}, which performed spectral energy distribution (SED) fitting on the SDSS galaxies using the Flexible Stellar Population Synthesis (FSPS) code \citep{Conroy2010}. We selected galaxies in the eFEDS field that lay between $0.01<z<0.1$. If a galaxy is closer, its emission may be individually resolved, and it may dominate the stacked signal. If it is farther, its emission may be too weak to contribute to the signal any more than the background noise. At these redshifts and given the \textit{eROSITA} angular resolution, we can spatially resolve the CGM. This produces a catalog of 2643 galaxies, whose properties are shown in Figure \ref{fig:sdss} and in the Appendix in Figure \ref{fig:hists}.

\subsection{\textit{eROSITA} data reduction}
In this analysis, we rely on the exposure-corrected eFEDS map in the $0.5-2.0$~keV energy range. The image is produced using the eFEDS event files and the \texttt{eSASS} data reduction package\footnote{\url{https://erosita.mpe.mpg.de/edr/}}. The eFEDS field is divided into four sections, each of which has a separate event list. Each event list is converted into an image with \texttt{evtool}. The corresponding exposure maps are created using the \texttt{expmap} command. Because of the light leaks in detectors T5 and T7 \citep{Predehl2021}, we exclude data from these two detectors. The images are projected onto the same WCS coordinates, and co-added using \texttt{dmimgcalc} task of  \texttt{CIAO} \citep{Fruscione2006}. 

Next, we use the eFEDS group and cluster catalog \citep{Liu2021} to exclude extended sources. Only 37 of the $>$2600 galaxies lie within a projected distance of $2R_{500}$ from a group or cluster. However, due to the much higher luminosity of the intracluster/group medium, emission from these objects may dominate the stack. The \citet{Liu2021} catalog provides the radius of maximum SNR, $R_{\rm max}$ for each object. Using scaling relations for the X-ray flux to determine $R_{\rm 500}$, we find that $0.3<R_{\rm max}/R_{\rm 500}<1$ for most objects. To be conservative, we mask out regions with $1.5R_{\rm max}$ around each extended source, and remove all galaxies in these regions from the stacks. We check that changing the masking radius to 1 or 2.5 $R_{max}$ does not change our results. 

Similarly, we mask out regions around resolved point sources, mostly distant AGN \citep{Salvato2021}. To this end, we use the 1.5 times the \texttt{APE-RADIUS}, which is the instrumental point spread function (PSF) at the position of the source, and is $\sim0.33'$. Increasing (decreasing) the masking radius by 50$\%$ decreases (increases) the surface brightness outside 20~kpc by $\sim10\%$.

We note that the masking applies to the photon counts, as well as the area and exposure maps. This ensures that galaxies hosting AGN, or partially overlapping with groups or clusters, are not down-weighted; each of their pixels either contributes to the counts, area and exposure maps or to none of them.

Finally, we extract the photons around each galaxy using the \texttt{CIAO} \texttt{dmextract} tool. Based on the stacked images in Figure \ref{fig:raw}, we expect a signal going out no further than 100~kpc, hence we produce radial profiles using annuli from $5.5-300$~kpc. We compute the background using the annuli at $150<r_{\rm kpc}<300$. Specifically, we first determine the average count rate within this annulus for each galaxy, then take the median count rate within each stack as the corresponding background. Thus, the background is computed for the stack as a whole. This removes sensitivity to contaminating, insufficiently masked point sources in individual background annuli. To further account for local variations, we compute the background at 150-300, 200-300 and 250-300~kpc, and use the range of these values to bracket the uncertainty in the background measurement.

To convert background-corrected counts to flux, we used an APEC model with a temperature of 0.5~keV, a metallicity of 0.3~$\rm{Z_\odot}$, and a column density of $N_H = 3\times10^{20} \ \rm{cm^{-2}}$. Given this spectrum and taking into account the \textit{eROSITA} effective area curves, the corresponding conversion factor is 1.624$\times10^{-12}$ erg/cm$^2$ \citep[e.g.][]{Brunner2021,Liu2021}. We use the luminosity distance $d_L$ at the redshift of each galaxy to convert the observed brightness to intrinsic luminosity. Although we compute radial profiles outside of 5.5~kpc, only use photons from 10-100~kpc around each galaxy when computing the total luminosity; this is because we expect to be dominated by the ISM and X-Ray Binaries (XRBs). 

\subsection{Contamination from XRBs and high-mass stars through the PSF}
\label{sec:xrbcont}
The emission from XRBs plays a major role in the overall X-ray appearance of normal galaxies \citep[e.g.][]{Gilfanov2004}. However, due to the relatively broad PSF of \textit{eROSITA}, a fraction of the counts associated with XRBs will contribute to the large-scale emission. Therefore, it is crucial to remove the contribution from XRBs when studying the CGM emission.

The emission from XRBs is often measured as a function of stellar mass and star formation rate; the former correlates with the low-mass XRBs (LMXBs) and the latter with high-mass XRBs (HMXBs). There are numerous studies of the XRB emission in galaxies, falling largely into two categories. Some studies \citep[e.g.][]{Lehmer2016,Aird2017,Fornasini2018} infer the XRB contribution from stacks of galaxies, splitting them by star-formation rate and stellar mass. While this probes a much larger sample of galaxies, these scaling relations are biased high since they are unable to separate the contribution from discrete point sources from the truly diffuse emission from the CGM and ISM. Studies of $\lesssim$100 nearby galaxies are able to resolve individual LMXBs and HMXBs \citep{Lehmer2010,Gilfanov2004,Mineo2012,Lehmer2019}. Since our goal is to isolate the CGM from XRBs, we utilize results obtained using the latter approach. In this work, we rely on the model derived by \citet{Lehmer2010}, which simultaneously characterizes the emission from LMXBs and HMXBs in the $0.5-2$~keV band. We note that the \citet{Lehmer2010} model is qualitatively similar to those obtained in other XRB population studies  \citep[e.g.][]{Gilfanov2004,Mineo2012}. However, the combined emission from XRBs is somewhat lower than that in \citet{Lehmer2019} model, which focused on low-metallicity galaxies containing a higher fraction of HMXBs.

Given the total XRB luminosity from \citet{Lehmer2010}, we convolve it with a PSF. To this end, we fit the \textit{eROSITA} PSF with a Moffat function \citep{Moffat1969}:
\begin{equation*}
    I (\theta) = I_0 \times [1+(\theta/\theta_s)^2]^\gamma
\end{equation*}
The best-fit model has a scale radius $\theta_{s}=9.40'$ and $\gamma=1.6$.
This allows us to account for the integrated contribution from PSF scattering between any two radii. For each stack, we renormalize the PSF such that the integrated luminosity is equal to that of the expected XRB contribution. To obtain the luminosity from the CGM, we subtract the normalized PSF integrated from $0<r_{kpc}<100$. For the surface brightness profiles, we subtract the PSF contribution in each radial bin.

A caveat in our analysis is that the stellar masses and star-formation rates are computed using the FSPS code, which utilizes SDSS spectra and photometry \citet{Montero2016}. While this approach results in accurate stellar masses, star-formation rates may have substantial uncertainties, which, however, likely cancel out in stacks of hundreds of galaxies. We note that results obtained with the FSPS code agree with other catalogs, such as MPA-JHU catalog, for quiescent galaxies; for the star-forming galaxies, however, the FSPS values on stellar masses are 30$\%$ lower and the star-formation rates are 80$\%$ higher than the MPA-JHU catalog. Using the MPA-JHU catalog would reduce the XRB contribution by 55$\%$ in the star-forming galaxy stacks; the contribution in quiescent galaxy stacks would remain unchanged. Acknowledging the uncertainty in measured properties of star-forming galaxies, we present two versions of each result: one removing only the LMXBs, and the other removing both LMXBs and HMXBs. We highlight, however, that the absolute magnitude of the star-formation rate does not affect our sorting of the galaxies, observed or simulated, into star-forming and quiescent; that is done simply by rank ordering.

\subsection{Simulated stacks}

\begin{figure}
    \includegraphics[width=0.45\textwidth]{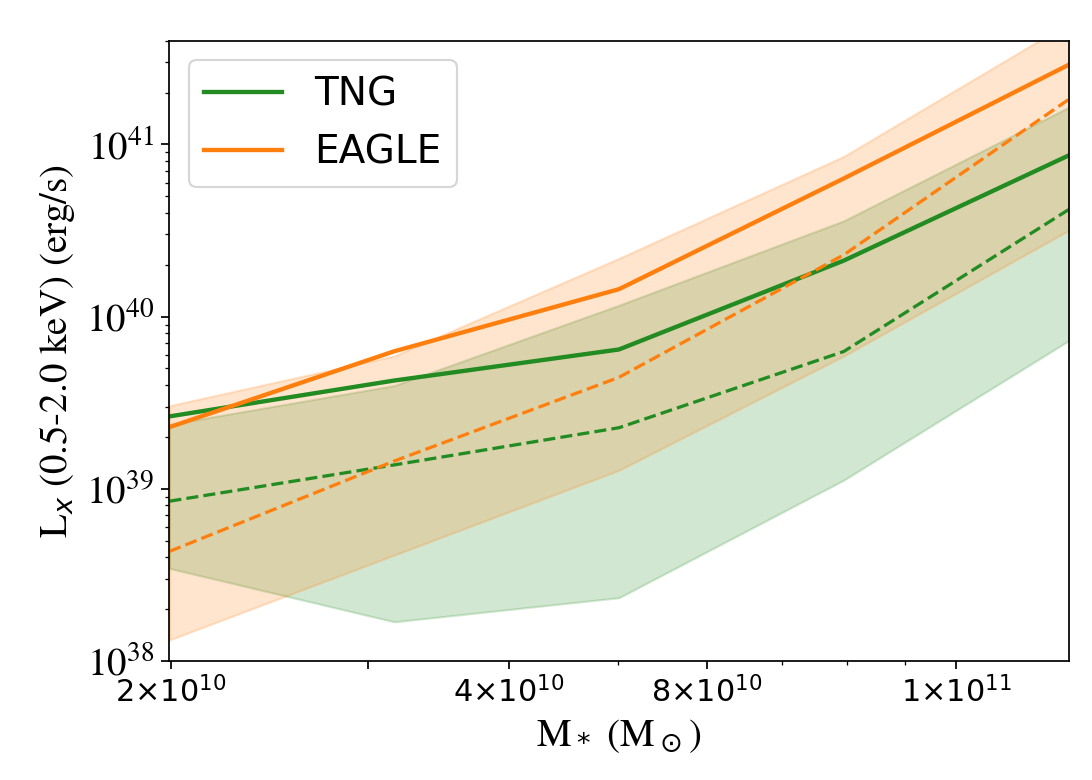}
    \caption{X-ray luminosity in the $0.5-2.0$~keV band, from $10-100$~kpc around each galaxy, for the simulated galaxies. EAGLE stacks are shown in orange, and TNG are shown in green. The solid line shows the mean at the given stellar mass, the dotted line shows the median, and the shaded regions show the $16-84$ percentile. The mean is $3-6$ times higher than the median at the low mass end, and two times higher at the high-mass end, and lies close to the 84th percentile.}
    \label{fig:Lx-Mstar-sim}
\end{figure}
 
\citet{Oppenheimer2020} suggested that the X-ray CGM will be observable by \textit{eROSITA} in stacks of simulated galaxies using two different simulations, EAGLE \citep{Schaye2015,Crain2015} and TNG100 \citep[][hereafter TNG]{Pillepich2018,Nelson2018}. In this paper, we re-sample the galaxies to match the observed stellar mass distribution in 0.1 dex $M_*$ bins, which makes the presented simulated predictions slightly different than \citet{Oppenheimer2020}. The resimulated stacks, matching the stellar mass to observations in bins of 0.1 dex, consist of 713 (287) and 963 (399) galaxies for the low (high)-mass stacks in EAGLE and TNG100, respectively. The galaxies in the top third by specific star formation rate (sSFR) are deemed star-forming and the bottom third is quiescent; the two sub-stacks have the same number of galaxies by construction. The simulated samples are compared to each other and the observations in the Appendix.

\begin{figure*}
    \centering
    \includegraphics[width=\textwidth]{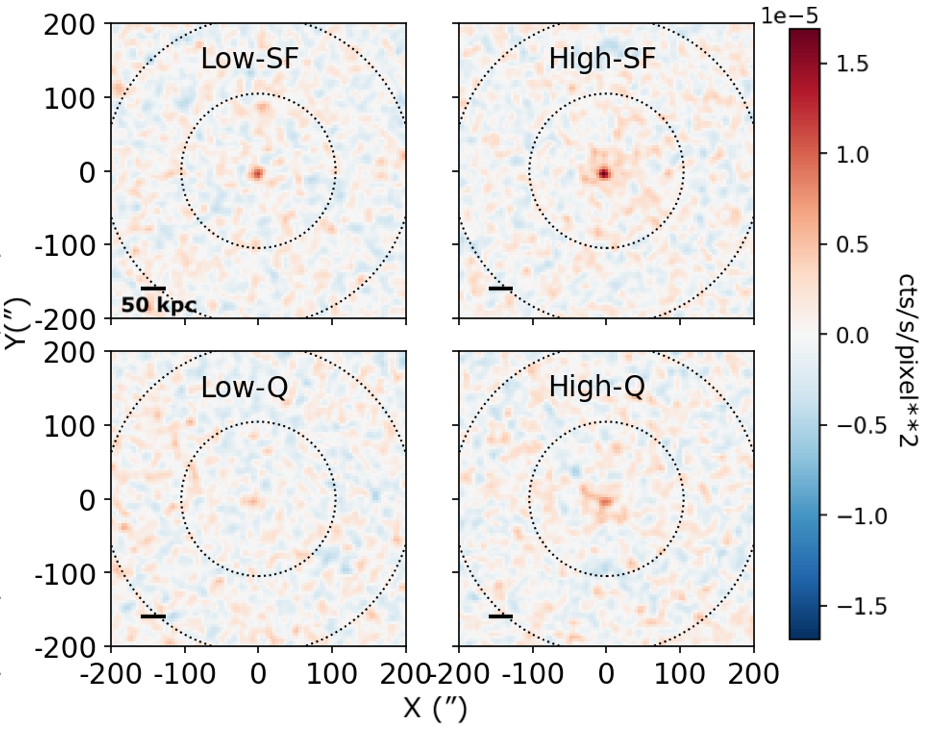}
    \caption{Background-corrected surface brightness maps of the star-forming (top) and quiescent (bottom) low-mass ($10.2<\log(M_*/M_\odot)<10.7$, left) and high-mass ($10.7<\log(M_*/M_\odot)<11.2$, right) stacks in the $0.5-2.0$~keV band. The annulus shows the region used to compute the background. Note that the emission from XRBs, scattered by the PSF, is not subtracted from these images. Emission is seen in all four stacks, but is more extended in the high-mass galaxies, and brighter in the center of star-forming galaxies. The scale bar corresponds to 50~kpc in each panel.}
    \label{fig:raw}
\end{figure*}

We briefly review our generation of the mock X-ray data from the simulations, detailed in \citet{Oppenheimer2020}.  We use the pyXSIM package\footnote{\url{http://hea-www.cfa.harvard.edu/~jzuhone/pyxsim/}} \citep{zuhone16} that generates mock photons via a Monte-Carlo sampling of X-ray spectra from the Astrophysical Plasma Emission Code \citep[{\sc APEC};][]{smith01} from each simulation fluid element with $T>10^{5.3}\;{\rm K}$ and hydrogen number density $ n_{\rm H} <0.22\;{\rm cm}^{-3}$.  Fluid elements belonging to the galaxy halo and out to at least three virial radii are included.  {\sc APEC} spectra assume collisional ionization equilibrium given the density, temperature, and the 9 individually-tracked element abundances in both EAGLE and TNG.  We mock Galactic foreground emission and absorption, Cosmic X-ray background sources, and instrumental noise from the SIXTE simulation software \citep{sixte}.  Mock projected observations of 2 ksec, applying the {\it eROSITA} PSF from SIXTE, are created assuming $z=0.01$, and stacked to match the sample characteristics described above.  For each galaxy field, we simulate a background field without the galactic haloes, and perform a subtraction to obtain a mock observation of the CGM source photons.  Although we do not mock the higher redshifts of the observed galaxies, we have experimented with higher redshift mocks to test that we recover surface brightness profiles at reduced spatial resolution.  We do not mock galactic sources, including XRBs, hot ISM, and AGN; therefore we subtract these three sources from the observational data.

Figure \ref{fig:Lx-Mstar-sim} shows the integrated $0.5-2$~keV band luminosity in the TNG (green) and EAGLE (orange) galaxies, computed from annuli at $10-100$~kpc around each galaxy. The mean luminosity of all the galaxies in a given mass bin is shown as a solid line, the median by the dashed line, and the $16-84$ percentiles are shown as shaded regions. The mean is closer to the median for the highest galaxy masses, and the ratio evolves faster in EAGLE than in TNG. The two simulations agree within 1-$\sigma$ uncertainties, with more similar results at lower stellar masses. We emphasize that stacking galaxies shows the mean signal, which is $1.5-5$ times higher than the median. This must be kept in mind when interpreting populations of galaxies based on stacked observations. 

\section{Results} 
\label{sec:results}
 We split the galaxies into two stacks by stellar mass, following \citet{Oppenheimer2020}. The low-mass bin has galaxies with $10.2<\log(M_*/M_\odot)<10.7$ and the high-mass bin has galaxies with $10.7<\log(M_*/M_\odot)<11.2$. Further, we sort each sample by specific star formation rate (sSFR); the top and bottoms thirds of galaxies are designated as star-forming and quiescent, respectively. 
 
Figure \ref{fig:raw} shows the stacked $0.5-2$~keV band X-ray maps in the two mass bins. There is a clear signal in both the low-mass and high-mass stacks with the emission being brighter and more extended in the high-mass galaxies. The star-forming galaxy stacks are brighter than the quiescent ones.

\begin{figure}
    \includegraphics[width=0.45\textwidth]{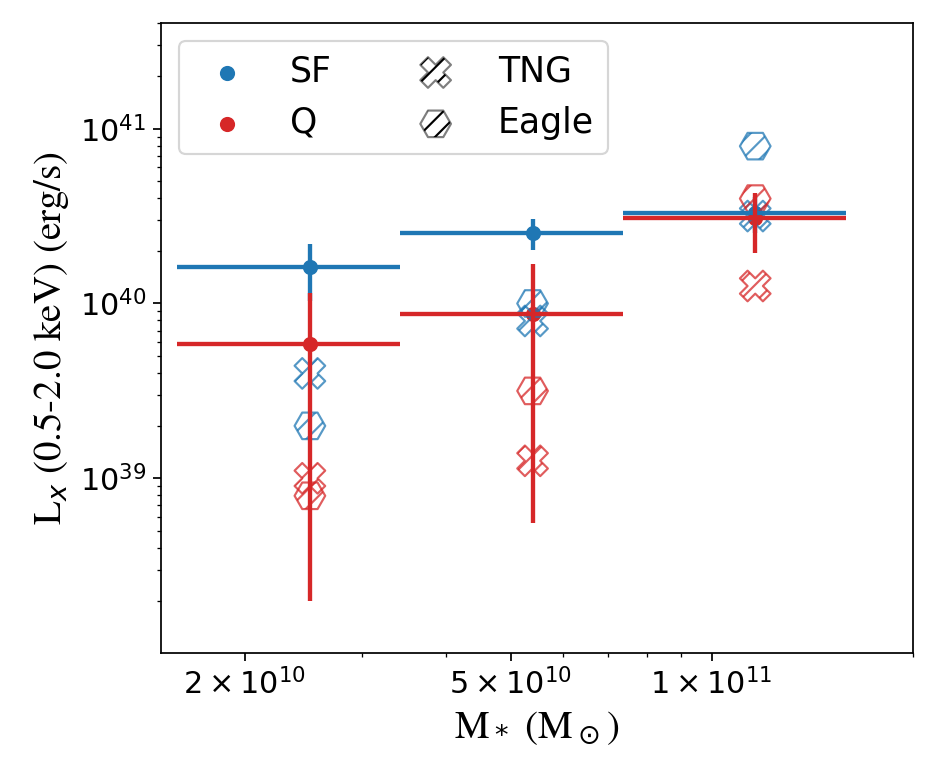}
    \caption{Comparison of the mean observed and simulated luminosity from $10-100$~kpc as a function of stellar mass, for star-forming (blue) and quiescent (red) galaxies, with the expected contribution from all XRBs removed. The observations are shown as crosses with 1-$\sigma$ Poisson error as well as systematic errors from changing the background level by 5$\%$. TNG is shown as crosses and EAGLE as hexagons.}
    \label{fig:Lx-Mstar-obs}
\end{figure}
 
Figure \ref{fig:Lx-Mstar-obs} compares the $0.5-2$~keV band mean luminosity of galaxies in an annulus of $10-100$~kpc in observations and simulations. The observed luminosities were corrected for XRB contamination. The luminosity of star-forming and quiescent galaxies are comparable, albeit star-forming galaxies are marginally more luminous. We checked that star-forming galaxies are brighter than quiescent galaxies if we exclude the central 20, 30 or even 40~kpc, so the difference cannot be explained by an underestimate of the central emission from galactic components such as XRBs. We emphasize that this difference is much less stark than in the simulations. A major difference between star-forming and quiescent galaxies is that the former have a flat luminosity function, while the latter grow brighter with stellar mass. However, the stellar mass dependence of quiescent galaxies is weaker than in the simulations.

\begin{figure*}
    \centering
    \includegraphics[width=\textwidth]{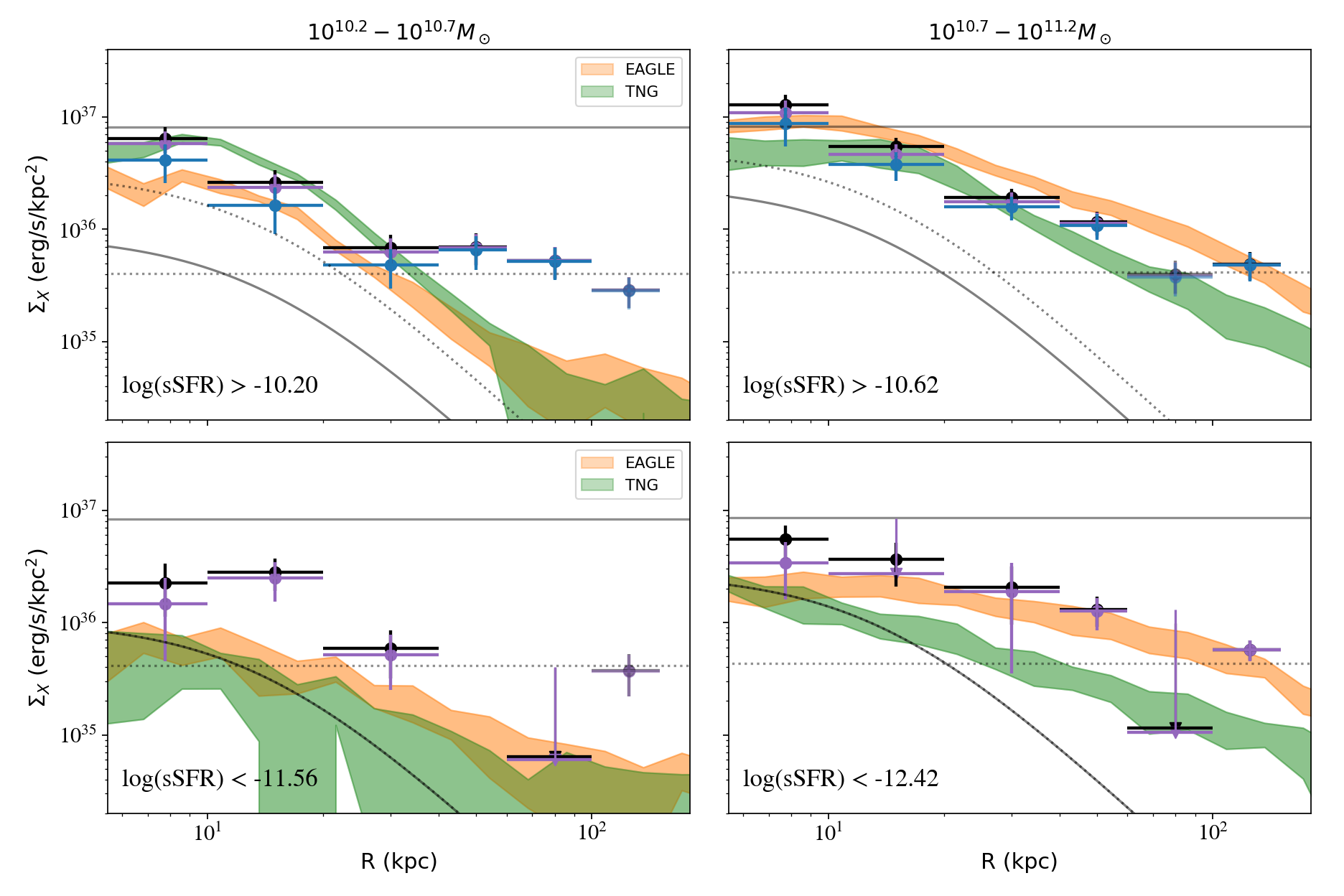}
    \caption{Surface brightness profiles of the low-mass (left) and high-mass (right) stacks, split into star-forming (top) and quiescent (bottom) thirds. The green bands show the predictions for the TNG simulations, while orange show EAGLE \citep{Oppenheimer2020}. The solid horizontal line shows the mean background for the stack, and the dotted line is 5$\%$ of this background. Black crosses show the background-corrected profile. Purple crosses remove the expected PSF-scattered contribution from low-mass X-ray binaries (LMXBs, shown as the solid curve), while blue additionally removes that from high-mass X-ray binaries and hot ISM (HMXBs, dotted curve). In the quiescent galaxies, by definition, the HMXB contribution is zero, so, we correct only for the LMXB. Where the Poisson error is greater than the mean, we show only the upper error bar and mark the mean with a downward triangle.}
    \label{fig:profiles}
\end{figure*}

Figure \ref{fig:profiles} shows the surface brightness profiles and compares them with  simulations. The purple crosses correct the observed flux for the contribution from LMXBs convolved with the PSF, shown as the solid black curve. The blue crosses additionally remove the contribution from HMXBs, shown as the dotted line. The latter contribution is negligible for the quiescent galaxy stacks, and is therefore not shown there. We emphasize that the XRB contamination has not been calibrated to the X-ray luminosities of the stacks themselves; rather, they only use the known galaxy properties and previously measured scaling relations. The solid horizontal line shows the mean background for the stack, where the background for each stack is the median surface brightness between $150<r_{\rm kpc}<300$. The dotted line shows $5\%$ of this background, below which systematic uncertainties associated with the background subtraction start to dominate. 

High-mass galaxies are brighter than those of lower mass, and star-forming galaxies are somewhat brighter than their quiescent counterparts. Star-forming galaxies have flatter profiles than the quiescent ones, particularly in the low-mass stack. The bright emission in the $40-100$~kpc region of low-mass star-forming galaxies is surprising and substantially exceeds theoretical predictions. We checked whether this signal is due to a few outliers, which are potentially contaminated by other sources along the line of sight. To this end, we sorted galaxies in the stack by their emission and removed up to a third of the most luminous ones. While this reduced the overall signal-to-noise, the signal outside 40~kpc persisted with small uncertainties. This suggests that the emission between $40-100$~kpc is real and could be of physical origin. Possibly, if individual star-forming galaxies expel blobs of gas, each in a different direction, the resulting average emission from these blobs may appear as spherically symmetric extended emission. This signal will be important to confirm with the full eRASS sample.

In the low-mass, quiescent galaxy stack, even though there is little ongoing supernova or AGN feedback, the simulated CGM is an order of magnitude dimmer than observations. The low-mass, star-forming galaxy stack agrees with EAGLE within 40~kpc, but is fainter than TNG; outside this region, it is an order of magnitude brighter than predicted by either simulation, with the net result being a very flat profile. The high-mass, quiescent galaxy stack agrees with EAGLE at $10-60$~kpc, but is brighter in the core and about five times dimmer further out; conversely, it is brighter than TNG within 60~kpc but agrees outside. The high-mass, star-forming galaxy stack overlaps with EAGLE, but is a factor of a few brighter than TNG at all radii.

\section{Discussion and Conclusions}
\subsection{New constraints for feedback prescriptions}
\label{sec:theory}

The comparison of simulated and observed profiles in Figure \ref{fig:profiles} offer insights into the roles of stellar and AGN feedback prescriptions. The quiescent galaxies, which tend to host more massive black holes and experience little ongoing feedback from star-formation processes, exhibit a cleaner signal of how AGN feedback operates over cosmic history. Similar to observations, both TNG and EAGLE show weaker X-ray luminosity in quiescent than star-forming galaxies, suggesting that the assembly of more massive black holes leads to greater integrated AGN feedback and the removal of more CGM gas to radii beyond 100~kpc \citep{Davies2019}. The large uncertainties for the low-mass quenched stacks in Figure \ref{fig:Lx-Mstar-obs} are consistent with both EAGLE and TNG, but the highest mass bin (with the smallest error bars) prefers EAGLE. The TNG quenched galaxies are more similar to observations at higher stellar masses, possibly reflecting that TNG is calibrated to reproduce baryon fractions in more massive group/cluster halos \citep{Pillepich2018}. At lower stellar masses, the simulations are several times fainter than the mean of the observations, although this needs to be confirmed with the improved signal-to-noise of the upcoming eRASS1 survey. This could suggest that the simulations rely too heavily on CGM ejection to quench their galaxies; alternatively, since the hot CGM produces primarily line emission, it may mean that the simulated CGMs are not sufficiently chemically enriched.

The surface brightness profiles of the high-mass, quenched galaxies agree well with EAGLE and are far brighter than TNG. EAGLE applies a single-mode, thermal AGN feedback model that becomes less efficient at removing the hot gas toward higher halo masses \citep{Davies2019}. Note that in EAGLE the group and cluster halos are also too baryon rich and X-ray luminous compared to observations \citep{Schaye2015}, so the high-mass bin here is possibly in a sweet spot where the EAGLE feedback prescription optimally matches observations. TNG applies a dual-mode AGN feedback prescription that is less mechanically efficient for high accretion rates, but pivots to an efficient jet-like kinetic mode at low-accretion rates that takes over above black hole masses of $\approx10^{8} \ \rm{M_\odot}$ \citep{Terrazas2020,Truong2021a}. In practice, almost all quenched galaxies in TNG have a black hole in the kinetic mode \citep{Piotrowska2022}. We speculate that the kinetic prescription may need to be recalibrated for the lower halo masses probed here, or it may need a novel dependence on halo or stellar mass.


The CGM of star-forming galaxies is likely more affected by recent stellar feedback prescriptions, particularly those associated with Type II supernovae and winds from massive stars. This feedback is weaker in TNG than EAGLE, resulting in a higher CGM baryon content in star-forming galaxies below $M_\star=10^{10.5} \rm{M_\odot}$ \citep{Davies2020}, and the correspondingly higher X-ray surface brightness for this simulation at lower stellar masses. While both EAGLE and TNG appear to show good agreement with the high-mass bin in Fig \ref{fig:profiles}, both fail to produce the flat extend emission observed outside 40~kpc in the low-mass stack; accordingly, the trend of luminosity with stellar mass in Figure \ref{fig:Lx-Mstar-obs} is steeper in the simulations than observed. While speculative, our observations support the intriguing possibility of stellar-driven superwinds more efficiently clearing the inner CGM and emitting more luminosity at larger radii. Our result could therefore have profound implications for the baryon cycle of the gas feeding galaxy assembly.

\subsection{Potential contamination from unresolved point sources}

In our analysis we mask out resolved point sources, including the central black holes and off-center ultra-luminous X-ray sources \citep{Swartz2004}. However, when stacking hundreds of galaxies, we may also be accumulating signal from unresolved central and off-center sources. Although some studies fit the black hole accretion rate as a function of star formation rate and stellar mass \citep{Aird2019,Carraro2020}, these models have large uncertainties for relatively low luminosity sources, such as those that remain unresolved in the eFEDS field. In fact, these studies suggest more luminous central emission in our stacks than the observed cumulative signal. Conservatively, we could assume that all emission at the center of the galaxies originates from point sources and convolve that empirical value with the point spread function. However, Figure \ref{fig:profiles} shows that the PSF has a different shape  than our measured profiles, implying that central point sources cannot explain all of the emission. The effect of off-center, unresolved ULXs and wandering black holes \citep{Ricarte2021} is beyond the scope of this paper.

\subsection{The importance of masking out extended structures}

\citet{Comparat2022} recently performed a similar study, stacking emission around $\gtrsim$16,000 galaxies from the GAMA sample at $0.05<z<0.3$. They repeated the stacking analysis with three different masks. The first, ``ALL", is identical to ours. The one focused on in their paper, however, is ``M1", which did not exclude groups/clusters in this redshift range. Crucially, this means including brightest group/cluster galaxies - which are usually classified as quiescent - that are surrounded by individually detectable X-ray emitting halos in their ``M1'' sample. This study concluded that quiescent galaxies are significantly more X-ray luminous than star-forming ones, in direct contrast to our results. 

We performed our analysis using a mask, which is identical with the ``M1'' mask of \citet{Comparat2022}. For our analysis, this corresponds to retaining 23 groups and clusters which are at $0.01 < z < 0.1$. By including these groups/clusters, the signal associated with the quiescent galaxy stacks were boosted, while the star-forming galaxy stack remained virtually invariable. Moreover, the quiescent galaxy stack grew brighter as we increased our radius of integration. Specifically, using the 0 $< r_{kpc} < 300$ region identical with \cite{Comparat2022}, we also find that the quiescent galaxy stacks are significantly brighter and their luminosity exceeds that of star-forming galaxies. Therefore, as was also concluded in that work, the bulk of the emission attributed to the CGM around quiescent massive galaxies in \citet{Comparat2022}, does not originate from the CGM, but is dominated by large-scale group/cluster atmospheres, which systems either host the galaxies or are projected at their position.

A closer comparison would be their ``ALL'' stacks, which appear consistent with our results. However, their galaxies are on average of $\sim3$ times more distant, making it difficult to spatially resolve the CGM. Because of the lower spatial resolution they also cannot exclude the central 10~kpc; therefore, their total luminosities cannot directly be compared to ours. 

\subsection{Conclusions}
In this work, we measured the large-scale X-ray CGM in stacks of 2643 galaxies observed in the \textit{eROSITA} eFEDS survey. Our results are: 
\begin{itemize}
    \item Even after masking out all identified extended and point sources, we detect emission from the CGM in galaxies of all stellar masses and star-formation rates.
    \item The X-ray luminosity of the CGM increases as a function of mass for quiescent galaxies, but its stellar mass dependence is much weaker than predicted in EAGLE and qualitatively different than in TNG. 
    \item In star-forming galaxies, the luminosity of the CGM is nearly identical in all stellar mass bins, in stark contrast with both TNG and EAGLE. This may imply that supernova feedback ejects the gas to significantly larger radii than predicted in models. 
    \item Star-forming galaxies are marginally brighter than their quiescent counterparts, but this difference is less significant than in TNG and EAGLE. 
    \item Our results suggest the need for more efficient supernova but less efficient AGN feedback models; they also indicate that quenching of galaxies cannot rely too heavily on ejecting the CGM. Part of the discrepancy may also be resolved by altering models for the chemical enrichment of the CGM, which also affects its emissivity at the energies probed in our study. 
   
\end{itemize}  
We detect the extended hot, X-ray emitting CGM in stacks of typical star-forming and quenched galaxies and are able to differentiate between the radial profiles of stacks of galaxies with different properties. The presence and characterization of the $\geq 10^6$ K CGM have enormous potential to answer outstanding questions brought about by ultraviolet surveys. The spatial resolution of our observations, along with the careful removal of contamination from clusters, groups, and AGN, allow us to disentangle the signatures of stellar and AGN feedback in a way that was not feasible in earlier studies. Upcoming simulations of galaxy evolution, which thus far have focused on reproducing optical properties of galaxies, should aim to alleviate the discrepancies with our newly observed X-ray profiles. This could mean modifying feedback physics alone; additionally, since metal-line emission likely dominates X-ray CGM emission, the solution may also involve altering the models of chemical enrichment of the CGM. The spatial resolution and the signal-to-noise ratio at larger radii will improve significantly with data for the full eRASS survey in the coming years.

\bigskip
\begin{small}
\noindent
\textit{Acknowledgements.} We thank the anonymous referee for helpful suggestions, and are grateful for discussions with Andy Goulding, Gerritt Schellenberger, Vinay Kashyap, James Davies and Charlie Conroy. R.P.K., \'A.B., C.J., and W.R.F. acknowledge support from the Smithsonian Institution and the Chandra High Resolution Camera Project through NASA contract NAS8-03060.  W.R.F. also acknowledges support from NASA Grants 80NSSC19K0116, GO1-22132X, and GO9-20109X. U. C. was supported by a Smithsonian Scholarly Studies Award. This study made use of high-performance computing facilities at Liverpool John Moores University.

This work is based on data from eROSITA, the soft X-ray instrument aboard SRG, a joint Russian-German science mission supported by the Russian Space Agency (Roskosmos), in the interests of the Russian Academy of Sciences represented by its Space Research Institute (IKI), and the Deutsches Zentrum für Luft- und Raumfahrt (DLR). The SRG spacecraft was built by Lavochkin Association (NPOL) and its subcontractors, and is operated by NPOL with support from the Max Planck Institute for Extraterrestrial Physics (MPE). The development and construction of the eROSITA X-ray instrument was led by MPE, with contributions from the Dr.\ Karl Remeis Observatory Bamberg \& ECAP (FAU Erlangen-Nuernberg), the University of Hamburg Observatory, the Leibniz Institute for Astrophysics Potsdam (AIP), and the Institute for Astronomy and Astrophysics of the University of T\"ubingen, with the support of DLR and the Max Planck Society. The Argelander Institute for Astronomy of the University of Bonn and the Ludwig Maximilians Universit\"at Munich also participated in the science preparation for eROSITA.
\end{small}

\bibliographystyle{aasjournal}
\bibliography{reference.bib}
\appendix

\begin{figure*}[ht!]
    \centering
    \includegraphics[width=\textwidth]{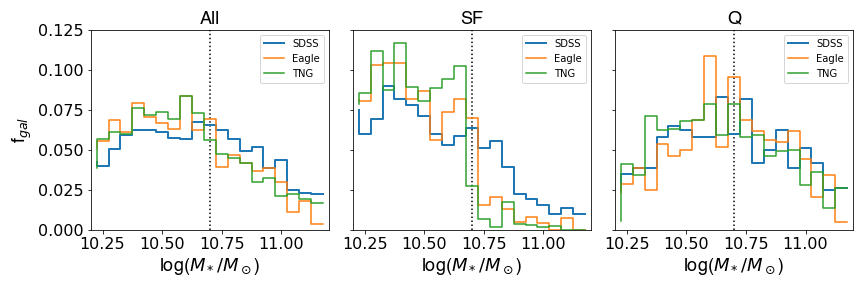}
    \includegraphics[width=\textwidth]{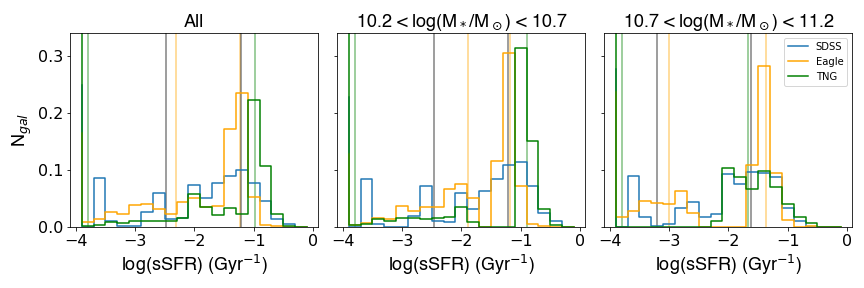}
    \caption{Comparison of stellar mass (top) and specific star formation rate (bottom) distributions in our observed sample (blue), EAGLE (orange) and TNG (green). The stellar mass distributions are additionally shown for the star-forming and quiescent bins, and the sSFRs for the low- and high-mass bins. The vertical lines indicate the cutoffs for the star-forming (SF) and quiescent (Q) substacks for the sample of the corresponding color.}
    \label{fig:hists}
\end{figure*}
TNG, EAGLE and SDSS have slightly different distributions of stellar masses and star-formation rates, shown in Fig \ref{fig:hists}. The simulations both aim to reproduce the stellar mass function at z = 0, and are successful to within $\sim 0.15$ dex. The simulations are not explicitly calibrated to reproduce SFR, although general agreement is achieved by broadly matching other observables. The limits for sSFR are somewhat different between the simulations and SDSS; however we note that these differences are comparable, if not slightly less than, the uncertainty in the observed star formation rates. The two simulations disagree with each other most significantly at the upper limit for low-mass quiescent galaxies, while the SDSS limit is in between. In other words, TNG is more aggressive in its quenching than observed galaxies, while EAGLE is gentler. Hence, rather than using absolute cuts in SFR or sSFR, we perform the same exercise of \citet{Oppenheimer2020} of dividing the galaxies up into the lower and upper third of sSFR in a stellar mass bin.
\end{document}